\documentclass[review,authoryear]{elsarticle}
\usepackage[english]{babel}
\usepackage{amsmath}
\usepackage{amsfonts}
\usepackage{amssymb}
\usepackage{bbold}
\usepackage{graphicx}
\usepackage{pict2e}
\usepackage{hyperref}
\usepackage{natbib}
\usepackage{color}
\usepackage[normalem]{ulem}
\usepackage{subfigure}
\usepackage{lineno}
\journal{Solar Energy}

\begin{document}

\begin{frontmatter}

\title{Comparison of statistical post-processing methods for probabilistic NWP forecasts of solar radiation}

\author{Kilian Bakker}
\author{Kirien Whan}
\author{Wouter Knap}
\author{Maurice Schmeits \corref{cor1}}
\ead{maurice.schmeits@knmi.nl}

\cortext[cor1]{Corresponding author}

\address{Royal Netherlands Meteorological Institute (KNMI), P.O. Box 201, 3730AE De Bilt, the Netherlands}

\begin{abstract}
    The increased usage of solar energy places additional importance on forecasts of solar radiation. Solar panel power production is primarily driven by the amount of solar radiation and it is therefore important to have accurate forecasts of solar radiation. Accurate forecasts that also give information on the forecast uncertainties can help users of solar energy to make better solar radiation based decisions related to the stability of the electrical grid. To achieve this, we apply statistical post-processing techniques that determine relationships between observations of global radiation (made within the KNMI network of automatic weather stations in the Netherlands) and forecasts of various meteorological variables from the numerical weather prediction (NWP) model HARMONIE-AROME (HA) and the atmospheric composition model CAMS. Those relationships are used to produce probabilistic forecasts of global radiation. We compare 7 different statistical post-processing methods, consisting of two parametric and five non-parametric methods. We find that all methods are able to generate probabilistic forecasts that improve the raw global radiation forecast from HA according to the root mean squared error (on the median) and the potential economic value. Additionally, we show how important the predictors are in the different regression methods. We also compare the regression methods using various probabilistic scoring metrics, namely the continuous ranked probability skill score, the Brier skill score and reliability diagrams. We find that quantile regression and generalized random forests generally perform best. In (near) clear sky conditions the non-parametric methods have more skill than the parametric ones.
\end{abstract}


\end{frontmatter}

\section{Introduction}{\label{sec:intro}}

Forecasts of meteorological variables are produced by projecting an initial state of the atmosphere into the future using a numerical weather prediction (NWP) model. In the projection there are unavoidable errors related to uncertainty in the initial condition and the physical parameterizations. This leads to uncertainty in the forecasts that can be quantified by generating a probabilistic forecast. In the case of solar radiation, this gives useful information about the uncertainty in power production of solar panels, as it is almost directly related to the uncertainty in forecasts of solar radiation. This helps users of solar energy make better solar radiation based decisions related to the stability of the electrical grid. \newline 
A considerable amount of work has already been done on generating probabilistic forecasts of global radiation (i.e. the total short-wave radiation on a horizontal surface received from a solid angle of 2$\pi$ steradians (unit: W/m2)). For example \citet{Lorenz_2009} produced 95\% confidence intervals on errors in the clear sky index (CSI, defined as global radiation divided by clear sky radiation) by fitting a polynomial function to the standard deviation of the errors as a function of the forecast CSI and the solar zenith angle. \citet{Verzijlbergh_2015} fitted a normal distribution to the CSI errors using NWP output from the GFS model of NCEP. Other distributions were also used for forecasting solar radiation, such as the beta and two sided power distribution in \citet{Fatemi_2018} and the gamma distribution in \citet{Bracale_2013}. They both use observations of various meteorological variables to forecast radiation. \newline
\citet{Massidda_2018} produced probabilistic forecasts by fitting quantiles of the underlying distribution of radiation, called quantile regression, using deterministic and probabilistic NWP output of ECMWF. They also extended this to a method called gradient boosting decision trees. \citet{Almeida_2015} use quantile regression forests to forecast quantiles of the distribution of solar power using NWP output from the WRF model run at Meteogalicia. Neural networks were used by \citet{Galvan_2017} to forecast confidence intervals of solar radiation using NWP output from the GEFS model of NCEP. \newline
In recent literature, there are some reviews that give an overview of papers using post-processing methods to produce probabilistic forecasts of solar radiation or solar power production. For example \citet{Antonanzas_2016} discussed papers using neural networks, random forests, support vector machines and nearest neighbours for forecasting solar power. \citet{Meer_2018} reviewed papers using quantile regression, quantile regression forests, Gaussian processes, bootstrapping, lower upper bound estimate, gradient boosting, kernel density estimation, nearest neighbours and an analog ensemble for forecasting solar power. \citet{Voyant_2017} discussed papers using linear regression, generalized linear models, neural networks, support vector machines, decision tree learning, nearest neighbours and Markov chains for forecasting solar radiation. This last review is oriented towards deterministic forecasts, whereas the first two focus on probabilistic forecasts. \newline
There are relatively few papers in which a comparison among multiple regression methods on the same data set for solar radiation or solar power forecasting is a primary goal. \citet{David_2018} made a comparison between probabilistic forecasts using random forests, neural networks, (weighted) quantile regression, gradient boosting decision trees, recursive GARCH and the sieve bootstrap. They used only past observations of global radiation and the CSI as predictors. \citet{Mohammed_2016} made a comparison between deterministic forecasts using decision trees, nearest neighbours, gradient boosting decision trees, random forests and lasso and ridge regression. They used ECMWF output variables as predictors in the regression methods. \citet{Zamo_2014} made a comparison between probabilistic forecasts using quantile regression and quantile regression forests with NWP output from the PEARP model at M\'et\'eo France as predictors. Lastly, \citet{Voyant_2018} made a comparison between probabilistic forecasts using quantile regression forests and gradient boosting decision trees with observed CSI as the only predictor. \\

In this paper we generate probabilistic forecasts of global radiation by applying a statistical post-processing technique called model output statistics (MOS, \citet{Glahn_1972}). This consists of fitting relationships between various NWP variables and CSI observations and using the relationships for producing skillful and reliable probabilistic forecasts. We make a comparison of probabilistic forecasts using an extensive list of regression methods, consisting of both parametric methods (gamma and truncated normal distribution) and non-parametric methods. The non-parametric methods consist of quantile regression, quantile regression forests, gradient boosting decision trees, and the relatively new generalized random forests and quantile regression neural networks. The methods are chosen to include the most commonly used methods and some (relatively) new methods, that take both linear and non-linear relationships into account. To our knowledge we are the first to make such an extensive comparison of probabilistic forecasting methods in a MOS setting. We use a large number of potential predictors, consisting of output from the high-resolution non-hydrostatic NWP model HARMONIE-AROME (HA, running at KNMI) and the atmospheric composition model CAMS (Copernicus Atmosphere Monitoring Service, running at ECMWF). This is a second novel aspect of this work, as previous work has not explored the benefit of including both NWP (here HA) and CAMS predictors, to our knowledge. We investigate the importance of the predictors in the fitted relationships. Additionally, the probabilistic forecasts are verified and compared with multiple scoring metrics to see which method performs best in which situation. These include the continuous ranked probability skill score, Brier skill score and reliability diagrams (e.g. \citet{Wilks_2011}). We also show the improvement with respect to the raw forecast (the global radiation forecast from HA) with the root mean squared error skill score and the potential economic value (e.g. \citet{Richardson_2000}). \\

The paper is structured as follows: in Section \ref{sec:data} we describe the research setting, consisting of the data, the regression methods, the fitting procedure and the scoring metrics. In Section \ref{sec:results} we present our results and discuss them in Section \ref{sec:discussion}. The paper is finished with conclusions in Section \ref{sec:conclusion}. 

\section{Data and Methods}{\label{sec:data}}

In this section we first describe the observations and model data that we used, including transformations applied to the model data. Next, we give an overview of the regression methods used to model the relationships between forecasts of weather variables (the predictors) and CSI observations (the predictand). The procedure we used for fitting is described and after that we describe verification measures to verify the methods. 

\subsection{Observations}

\begin{figure}[!ht]
\centering
    \includegraphics[width = 0.9\linewidth]{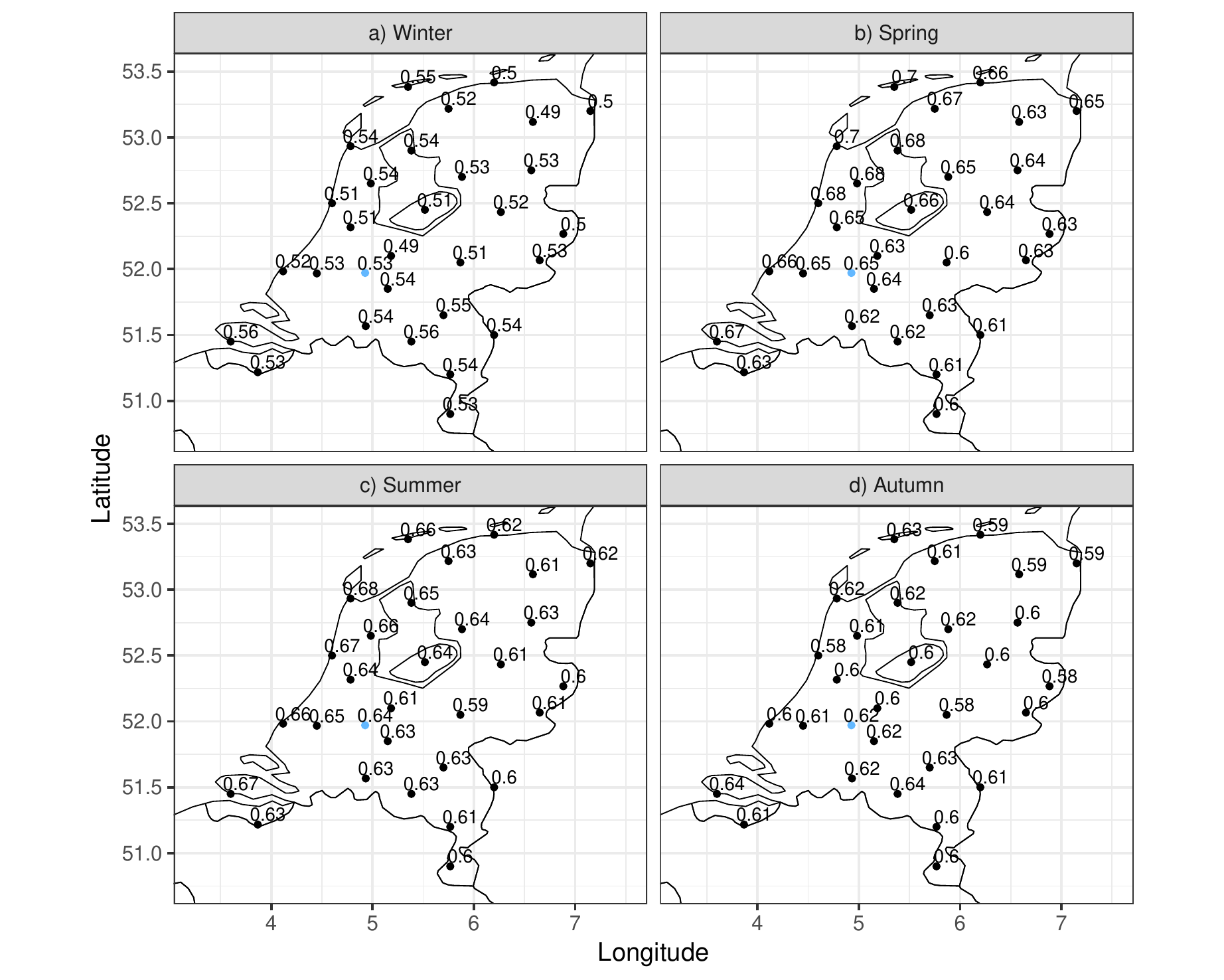}
    \caption{The KNMI station locations and seasonally averaged observed CSI in the Netherlands. Station Cabauw is marked in blue.}
    \label{fig:spatialCSIobs}
\end{figure}

The meteorological variable we study in this paper is the global radiation. We used hourly average observations of global radiation measured at 30 weather stations in the Netherlands by the KNMI network of automatic weather stations for the period of April 1, 2016 to March 16, 2018 (\citet{KNMI_2018}). The KNMI network of solar radiation measurements consists of unventilated Kipp \& Zonen CM11 pyranometers. The calibration of the instruments is traceable to the World Radiometric Reference (WRR). The instruments are sampled every 12s and 1 hr averages are used for the analysis presented here. Routine quality control procedures are applied based on intercomparison of stations and clear-sky comparisons. The achieved uncertainty is within the limits of “network operations” as defined by the WMO CIMO guide (8\% for hourly totals, 95\% confidence level; \citet{WMO_2017}). We transformed the global radiation into CSI by dividing them by the hourly average clear sky radiation. This removes the large seasonal and diurnal cycle of radiation. The clear sky radiation is calculated using the European Solar Radiation Atlas (ESRA) model (\citet{ESRA}), with as inputs the solar zenith angle (\citet{Mich_algo}) and the Linke turbidity factor (monthly average values for De Bilt are used from \citet{TL}). The locations of the stations and the seasonally averaged observed CSI are shown in Figure \ref{fig:spatialCSIobs}. For inland stations the seasonally averaged observed CSI is slightly lower than for coastal stations in spring and summer due to convection. In winter and autumn the pattern is quite homogeneous.

\begin{figure}[!ht]
    \centering
    \includegraphics[width = \linewidth]{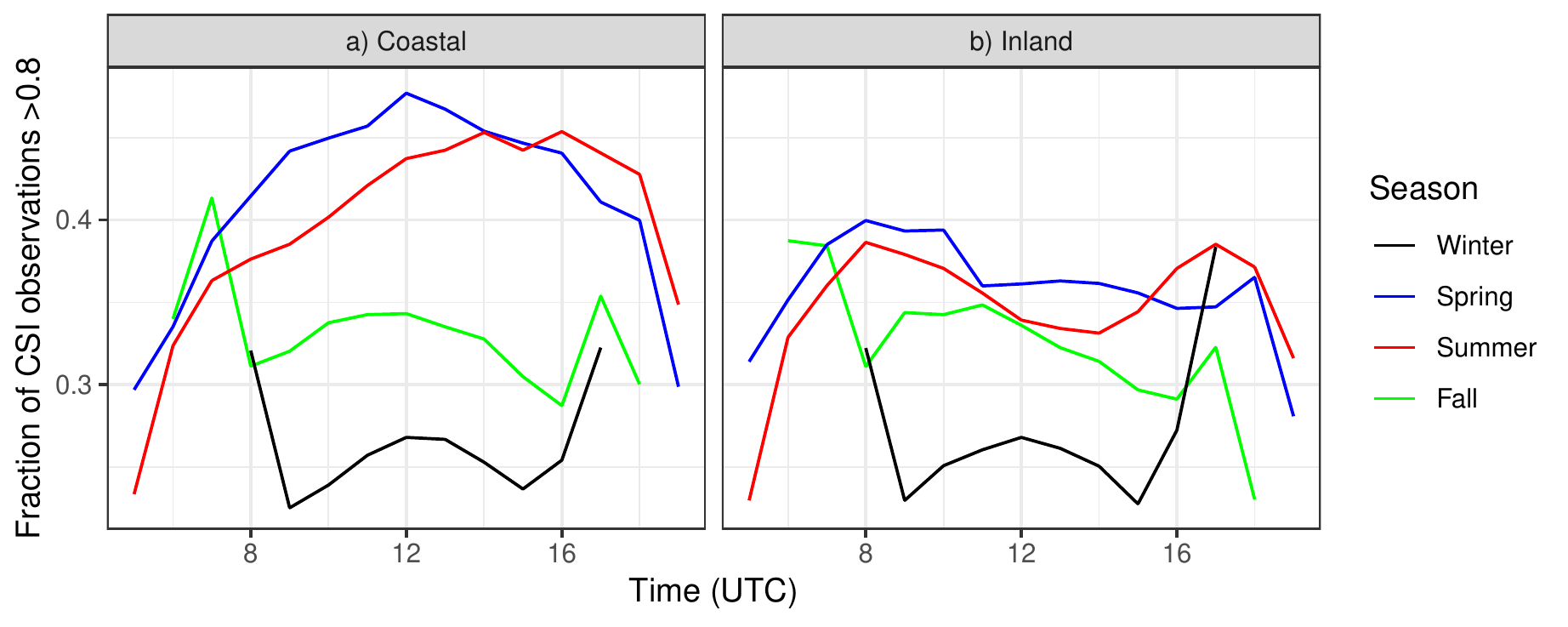}
    \caption{The fraction of CSI observations $> 0.8$ versus the time of the day for the four seasons and for (a) coastal (less than 25 km from the coast) and (b) inland (more than 25 km from the coast) stations.}
    \label{fig:timeseriesCSIobs}
\end{figure}

Figure \ref{fig:timeseriesCSIobs} shows daily courses of the fraction of CSI observations strictly greater than 0.8 (corresponding to low cloud cover conditions, including clear skies) for the four seasons and for coastal (less than 25 km from the coast) and inland (more than 25 km from the coast) stations separately. At the inland stations during spring, summer and fall the fraction drops during the day a few hours after sunrise. This is likely due to convection with a (secondary) maximum at the end of the day caused by the disappearance of shallow convection. Inland stations during winter and coastal stations in all seasons show a different diurnal cycle with a similar evolution between a few hours after sunrise and a few hours before sunset and a (secondary) maximum around 12 UTC. In spring and summer the fraction of low cloud cover events is higher at the coastal stations than at the inland stations due to less convective events. 

\subsection{Model data}

The model data that we used as the potential predictors for the regression methods consist of forecasts of atmospheric temperature, humidity and different cloud, radiation and aerosol variables (Table \ref{table:predictors}). Most of the predictors (marked with *) are NWP output of HA (\citet{Bengtsson_2017}). The aerosol predictors (marked with ** ) are forecasts produced by CAMS (\citet{Flemming_2017}). Because HA uses climatological averages for the aerosols, we expect the aerosol forecasts as predictors to have added value in fitting the relationships (see also \citet{Alfadda_2018}). \\
We used hourly output from HA for lead times 0-48 hours at the grid points closest to the stations (in the same date range as the observations, April 1, 2016 to March 16, 2018). The CAMS lead times are 3-hourly and therefore we take the lead time closest to the corresponding time of the observations. The CAMS forecasts are later available than HA and therefore we use them for lead times of 24-72 hours from the previous day. We used CAMS data for the same period as the HA and observation data. More information about HA and CAMS can be found in Table \ref{table:models}.

\begin{table}[!ht]
    \centering
    \begin{tabular}{|c|c|c|} \hline
    & HA & CAMS \\ \hline
    Version(s) & 38 & 43r3, 43r1, 41r1 \\ \hline
    Type of forecasts & \multicolumn{2}{|c|}{Deterministic} \\ \hline
    initial time & \multicolumn{2}{|c|}{0 UTC} \\ \hline   
    Time domain & 0-48 hours & 24-72 hours \\ \hline
    Time resolution & hourly & 3-hourly \\ \hline
    Spatial domain & Western Europe (800x800 grid points) & Global \\ \hline
    Spatial resolution & 2.5km x 2.5km & $0.5^{\circ}$ x $0.5^{\circ}$ \\ \hline
    Vertical levels & 65 & 60 \\ \hline
\end{tabular}
    \caption{Information about the models HA and CAMS.}
    \label{table:models}
\end{table}

\begin{table}[!ht]
\centering
\subfigure[HA = *, CAMS = **. 'surf' stands for the Earth's surface and 'toa' for the top of the atmosphere (defined as the height with air pressure equal to 0.01 hPa). The layers indicate the lower (0-2000 m), middle (2000-6000 m), higher (6000 m-TOA) and total (0 m-TOA) part of the atmosphere.]{
\begin{tabular}{|c|c|} \hline
    & Potential predictors \\ \hline
    Temperature* & $T_{\text{layers}}$ \\ \hline
    Relative humidity* & $RH_{\text{layers}}$ \\ \hline
    Radiation* & $G$, $DIR_{\text{surf}}$, $DIR_{\text{toa}}$, $NCS_{\text{surf}}$, $NCS_{\text{toa}}$ \\ \hline
    Clouds and rain* & $RAIN$, $CC_{\text{layers}}$, $CW_{\text{layers}}$, $PW_{\text{layers}}$ \\ \hline
    Aerosols** & $AOD$, $ANG$, $OZ$ \\ \hline
    Time/place & $LAT$, $LON$, $DOY$, $\cos{z}$, $DIST_{\text{coast}}$, $DIST_{\text{water}}$, $DIST_{\text{inland}}$ \\ \hline
    \end{tabular}
}
\subfigure[The abbreviations of the potential predictors explained. Net clear sky stands for the incoming minus the outgoing radiation under a clear sky.]{
    \begin{tabular}{|c|c|c|} \hline
    $T$ & Temperature \\ \hline
    $RH$ & Relative humidity \\ \hline
    $G$ & Global radiation \\ \hline
    $DIR$ & Direct radiation \\ \hline
    $NCS$ & Net clear sky (global) radiation \\ \hline
    $CC$ & Cloud cover \\ \hline
    $CW$ & Cloud water \\ \hline
    $PW$ & Precipitable water \\ \hline
    $AOD$ & Vertically integrated aerosol optical depth at 500 nm \\ \hline
    $ANG$ & Vertically integrated \r{A}ngstr\"{o}m exponent \\ \hline
    $OZ$ & Vertically integrated ozone \\ \hline
    $LAT$ & Latitude \\ \hline
    $LON$ & Longitude \\ \hline
    $DOY$ & Day of the year \\ \hline
    $DIST$ & Distance \\ \hline
    \end{tabular}
}
\caption{The potential predictors}
\label{table:predictors}
\end{table}

For each of the potential predictors with 'layers' as a subscript, we computed them for the lower (0-2000 m), middle (2000-6000 m) and upper atmosphere (6000 m and higher), and also an aggregated version of the total atmosphere (except for temperature and relative humidity, where the total atmosphere component is missing). The HA output consists of 65 levels in the atmosphere and for the cloud water we applied a weighted average (with weights equal to the distances between the levels) to reduce these to the four layers (lower, middle, upper and total atmosphere). For temperature and humidity we used HA output from 11 pressure levels and reduced these to the three layers (lower, middle and higher part of the atmosphere) by applying a weighted average. The precipitable water is computed using temperature and humidity from the 11 pressure levels together with the method outlined by \citet{PW} and applying a weighted average. \newline
We transformed all radiation predictors into hourly average CSI values in the same way as the observations. The rain predictor values are precipitation sums over each hour. For all other potential predictors the valid time is used. \newline
We also calculated time/place predictors, which are used as additional potential predictors. These consist of the latitude and longitude of the station locations, the day of the year, the distance to the coast (minimum Euclidean distance to the sea), distance to water (minimum Euclidean distance to the sea or one of the lakes), distance to inland (the Euclidean distance to the intersection point of the borders of the Netherlands, Belgium and Germany) and the cosine of the solar zenith angle.

\subsection{Regression methods}{\label{sec:regMethods}}

We generated probabilistic forecasts using regression methods that find relationships between the predictors and the predictand. The relationships help in understanding how the value of the predictand changes when any one of the predictor values change. Then, one can quantify the uncertainty in the predictand in terms of the predictors and form the probabilistic forecast. We fit for 7 different regression methods and they are described in more detail in the following subsections.

\subsubsection{Parametric regression}

Parametric regression methods assume a specific predictive distribution with parameters $\{P_1, P_2, P_3, P_4\} = \{\mu, \sigma, \tau, \nu\}$ (location, scale, skewness and kurtosis). The parameters are first chosen as constant values and then fitted sequentially (one by one, each new one using the already fitted ones) with linear relationships:
\begin{equation}{\label{eq:LR_dist}}
    P_i = \beta_0 + \beta_1 X_1 + ... + \beta_p X_p, i = 1,...,4,
\end{equation}
where $X_j, j = 1,...,p$, are the predictors and $\vec{\beta} = \{\beta_0,\beta_1,...,\beta_p\}$ are the regression parameters. The optimal linear relationship is found by minimizing the log-likelihood function (i.e. the negative logarithm of the forecast probability density function). \newline
We avoid overfitting by using only a selection of the predictors in the model. The selection that gives the best fit is found by forward (adding one predictor) and backward (removing one predictor) stepwise selection that aims to minimize the Akaike Information Criterion (AIC). The stepwise procedure is applied separately for each distribution parameter. \newline
We apply this regression method in the programming language R using the gamlss package (\citet{R_GAMLSS}) for two distributions: the gamma distribution (GA) and the truncated normal distribution (NOTR). The truncated normal distribution is formed by left truncation at zero of the normal distribution (\citet{GAMLSS}) and has therefore, just like the gamma distribution, only probability mass for positive values, which suits for forecasting radiation.

\subsubsection{Quantile regression}

Quantile regression fits a number of quantiles $q_k, k = 1,...,Q$ from the underlying distribution separately by assuming the linear relationship:
\begin{equation}{\label{eq:LR_quants}}
    q_k = \beta_0 + \beta_1 X_1 + ... + \beta_p X_p, k = 1,...,Q
\end{equation}
The optimal linear relationship is found by minimizing the weighted sum of residuals (WSR, defined as observations - fitted quantiles), with weights equal to $q_k$ and $q_k - 1$ for respectively positive and negative residuals (\citet{QR}). Regularization to prevent overfitting was not successful in our study and therefore we apply the stepwise procedure, this time with the aim of minimizing the average over the AIC values for the different quantiles. Consequently, all quantiles are fit with the same predictors. It occasionally happens that negative values are forecast and in that case we set them to zero. Also, we solve crossings between quantiles by sorting them in ascending order. Lastly, some small random noise, generated from a normal distribution with mean zero and a variance of 0.001, is added to the predictors to prevent the matrix with the predictors from becoming singular. Quantile regression (QR) is applied in R using the quantreg package (\citet{R_QR}).

\subsubsection{Quantile regression neural network}

Quantile regression neural networks are a non-linear extension of quantile regression. It can be visualized by an input layer (the predictors), one or more hidden layers and an output layer (the forecast for one quantile). In the hidden layers we adjust the input towards the output by multiplication with weights, adding some bias and applying a non-linear activation function, taken here as the default $\frac{1}{1 + \exp{(−x)}}$. The optimal weights and bias are found by minimizing the WSR in a boosting fashion where the negative gradient is followed for a number of iterations. More details can be found in \citet{QRNN}. \newline
Regularization to prevent overfitting was not successful and therefore we apply the stepwise procedure. Further, negative forecasts are set to zero. Crossings between quantiles are solved by adding the quantiles as a predictor in the input layer with a monotonically increasing constraint (\citet{MCQRNN}). This is called the monotone composite extension of quantile regression neural network (MCQRNN) and we apply it in R using the qrnn package (\citet{R_QRNN}).

\subsubsection{Random forests}

Random forests have binary regression trees (\citet{CART}) as building blocks. A binary regression tree is formed by splitting the data set into subsets by applying thresholds on the predictors. Each split is chosen to minimize the variance in the predictand. The splitting is repeated until some stopping criterion is reached, such as a minimum number of predictand values left in each subset. A random forest is a collection of trees (\citet{RF}) and reduces the chance for overfitting by constructing each tree on a different (bootstrap sampled) subset of the data. To make the trees even more independent, at each split only a (bootstrap sampled) subset of the predictor set is used. Probabilistic forecasts are made by looking at the underlying distribution of the predictand values in the terminal nodes of all trees and drawing quantiles from it (\citet{QRF}). This method is called quantile regression forests (QRF) and is applied in R using the quantregForest package (\citet{R_QRF}). \newline
Generalized random forests (GRF) is very similar to QRF. Each split is now chosen to maximize the difference between the underlying distributions of the predictand values (\citet{GRF}). Also, whereas QRF uses bootstrap sampling with replacement, GRF uses bootstrap sampling without replacement. GRF is applied in R using the grf package (\citet{R_GRF}).

\subsubsection{Gradient boosting decision trees}

Gradient boosting decision trees boosts the fitted quantiles (separately) towards the actual quantiles of the underlying distribution by following the negative gradient of the WSR for a number of iterations. The negative gradients are estimated using regression trees. Then in each iteration the fits are updated by adding the estimated negative gradients multiplied with the learning rate (\citet{GBM}). The learning rate is generally chosen to be small ($\approx 0.1$) to prevent overshooting the minimum in the WSR. To reduce the chance for overfitting, each tree is constructed on a (bootstrap sampled) subset of the data (\citet{SGBM}). Consequently, in each iteration we only improve the fitted quantile for the predictand values in the chosen subset. Further, negative values are set to zero and crossings between quantiles are solved by sorting them in ascending order. Gradient boosting decision trees (GBDT) is applied in R using the gbm package (\citet{R_GBM}). 

\subsection{Fitting procedure}

We fitted relationships between the potential predictors and the predictand using the 7 methods. The fitting was done per lead time, because systematic errors increase with lead time, and for all stations at once. We split the data set into a training (for model fitting) and testing (for verification) set. The testing set consists of 1/3 of the data set containing consecutive dates and the rest of the data are in the training set. This leads to around 3200 training and 1600 testing cases for each fit when we fit to each season separately. We applied a 3-fold cross-validation. For each lead time we only take combinations of dates and stations during daylight into account, defined as a clear sky radiation higher than 20 W/m$^2$. Only for lead times +5h till +19h and +29h till +43h there are enough (defined by at least 50) combinations of dates and stations during daylight to make the fit. For each method, we generated probabilistic forecasts consisting of 49 quantiles, distributed evenly between 0.02 and 0.98 with a gap of 0.02 between them. \newline 
We can either fit the methods on the whole year or on the four meteorological seasons separately, defined by the Winter (December, January, February), the Spring (March, April, May), the Summer (June, July, August) and the Autumn (September, October, November). As shown in \citet{Thesis_Kilian}, fitting for each season separately produces better fits and is therefore used in producing the results. Also averaging the potential predictors from HA over 3 lead times and over a 9x9 block of grid points gives better fits. Therefore, we apply both the temporal and spatial smoothing to our potential predictors. Next, the optimal hyper-parameter settings for each method can be found in \citet{Thesis_Kilian} (only the optimal settings for MCQRNN have changed) and are summarized in Table \ref{table:optimal_settings}.

\begin{table}[!ht]
\centering
    \begin{tabular}{|c|c|c|c|c|} \hline
     & GA & NOTR & QR & MCQRNN  \\ \hline \hline
    \#steps ($\mu$ or quantiles) & 5 & 5 & 5 & 5 \\ \hline
    \#steps ($\sigma$) & 1 & 1 & - & -  \\ \hline
    \#hidden layers & - &  - & - & 1 \\ \hline
    \#deep hidden layers &  - & - & - & 0 \\ \hline
    Iterations &  - & - & - & 10 \\ \hline
    Penalty &  - & - & - & 0 \\ \hline
    \end{tabular}
    \vspace{1cm}
    \begin{tabular}{|c|c|c|c|} \hline
     & QRF & GRF & GBDT \\ \hline \hline
    \#trees & 500 & 500 & 100 \\ \hline
    Minimal nodesize & 5 & 5 & 5  \\ \hline
    Case sampling fraction & 1/2 & 1/2 & 1/2 \\ \hline
    Predictor sampling fraction & 1/3 & 1/3 & - \\ \hline
    Depth of trees & - & - & 1 \\ \hline
    Learning rate & - & - & 0.1 \\ \hline
    \end{tabular}
\caption{The optimal hyper-parameter settings for each method. The descriptions of the hyper-parameters are found in \citet{Thesis_Kilian}.}
\label{table:optimal_settings}
\end{table}

\subsection{Scoring metrics}{\label{sec:SS}}

To quantify the performance of the regression methods, we calculated scoring metrics over the forecasts (e.g. \citet{Wilks_2011}). First, the medians of the forecasts are verified using the root mean squared error (RMSE):
\begin{equation}{\label{eq:RMSE}}
    \text{RMSE} = \sqrt{\frac{1}{n}\sum_{i=1}^{n} \limits (y_i - f_{i,\text{med}})^2},
\end{equation}
where $\vec{y} = (y_1,...,y_n)$ are the CSI observations (for the same season and (valid) time as the forecasts) and $F = \{f_{i,k}, i = 1,..,n, k = 1,...,Q\}$ the forecasts, with 'med' denoting the 0.5 quantile. The RMSE is transformed to a skill score using the RMSE of the sample climatology (the mean of the observations): 
\begin{equation}{\label{eq:SS}}
    \text{RMSE\_SS} = 1 - \frac{\text{RMSE}}{\text{RMSE}_{\text{clim}}}
\end{equation}
For verifying the probabilistic forecasts we used the continuous ranked probability score (CRPS):
\begin{equation}{\label{eq:CRPS}}
    \text{CRPS} = \frac{1}{n} \sum_{i=1}^{n} \limits \left( \frac{1}{Q}\sum_{k=1}^{Q}\limits |f_{i,k} - y_i| - \frac{1}{2Q^2}\sum_{k=1}^{Q}\limits \sum_{l=1}^{Q}\limits |f_{i,k} - f_{i,l}| \right)
\end{equation}
We calculated the continuous ranked probability skill score (CRPSS) using Equation (\ref{eq:SS}) with the RMSE replaced by the CRPS. The sample climatology consists in this case of quantiles drawn from the observations to generate a sample climatological distribution. \\
For verification it is also informative to look at probabilities of not exceeding a certain threshold $T$ of CSI. Therefore we have binary observations ($\tilde{y}_i = \mathbb{1}_{y_i \leq T}, i = 1,...,n$) and forecast probabilities ($\tilde{f}_i = \frac{1}{Q}\sum_{k=1}^{Q}\limits \mathbb{1}_{f_{i,k} \leq T}, i = 1,...,n$), which are verified using the Brier score (BS):
\begin{equation}{\label{eq:BS}}
   \text{BS} = \frac{1}{n} \sum_{i=1}^{n} \limits (\tilde{y}_i - \tilde{f}_i)^2 
\end{equation}
The skill is measured by the Brier skill score (BSS), calculated using Equation (\ref{eq:SS}) with the RMSE replaced by the BS, where the sample climatology is defined as the mean of the binary observations. \newline
We also produced reliability diagrams that compare forecast probabilities with observed relative frequencies. A perfect reliable forecast has a 1:1 correspondence between them. \newline
The potential economic value measures the potential economic impact of the forecasts and is defined by a simple cost-loss model described in \citet{Richardson_2000}. It looks at the event of not exceeding a certain threshold $T$ of CSI with the associated contingency table (Table \ref{table:PEV}). A hit and false alarm lead to some costs $C$, a miss to a loss $L$ and with a correct rejection there are no costs or losses. 

\begin{table}[!ht]
    \centering
    \begin{tabular}{|c|c|c|}\hline
     & Observed & Not observed \\ \hline
    Forecast & Hit $\rightarrow C$ & False alarm $\rightarrow C$ \\ \hline
    Not forecast & Miss $\rightarrow L$ & Correct rejection $\rightarrow 0$ \\ \hline
    \end{tabular}
    \caption{Contingency table corresponding to the event that the threshold of CSI does not exceed a certain value $T$. $C$ indicates costs and $L$ losses.}
    \label{table:PEV}
\end{table}

The potential economic value (PEV) is calculated as:
\begin{equation}{\label{eq:PEV}}
    \text{PEV} = \left\{\begin{matrix} \frac{\frac{C}{L}(H + FA - 1) + M}{\frac{C}{L}(\text{ORF} - 1)}  & \text{if } \frac{C}{L} < \text{ORF} \\ \frac{\frac{C}{L}(H + FA) + M - \text{ORF}}{(\frac{C}{L} - 1)\text{ORF}} & \text{otherwise} \end{matrix}\right.,
\end{equation}
where $H, FA$ and $M$ are the hit, false alarm and miss frequencies and ORF the observed relative frequency.

\section{Results}{\label{sec:results}}

In this section we first investigate how important the predictors are for the different methods. Subsequently, we show the performance of the methods according to all scoring metrics defined in Section \ref{sec:SS}. 

\subsection{Importance of predictors}

We have 34 potential predictors and some of them are more important than others. In Tables \ref{table:ranking_pred_stepwise} and \ref{table:ranking_pred_tree} we list the 10 most important predictors for the stepwise-based and tree-based methods, respectively. For the stepwise-based methods (GA, NOTR, QR and MCQRNN) the ranking is based on the amount of times the predictors are chosen in the procedure. For the tree-based methods (QRF, GRF and GBDT) the ranking is based on the improvement the predictors have on the fitted trees, considered over all splits where the corresponding predictor is chosen. This improvement is defined slightly differently in QRF, GRF and GBDT and more details can be found in \citet{R_RF}, \citet{R_GRF} and \citet{R_GBM}. The final ranking is then the average improvement over all trees in all fits. Due to the different definitions of the importance measure between the stepwise-based and tree-based methods, we cannot directly compare the resulting rankings between these methods. 

\begin{table}
    \centering
    \begin{tabular}{|c|c|c|c|c|} \hline
        Rank & GA & NOTR & QR & MCQRNN  \\ \hline
        1 & $G$ & $G$ & $RH_{\text{low}}$ & $G$ \\ \hline 
        2 & $RH_{\text{low}}$ & $RH_{\text{low}}$ & $G$ & $RH_{\text{low}}$ \\ \hline 
        3 & $DIR_{\text{surf}}$ & $RH_{\text{middle}}$ & $DIR_{\text{surf}}$ & $RH_{\text{middle}}$ \\ \hline 
        4 & $RH_{\text{middle}}$ & $DIR_{\text{surf}}$ & $CC_{\text{high}}$ & $CC_{\text{total}}$ \\ \hline 
        5 & $CC_{\text{total}}$ & $CC_{\text{total}}$ & $CC_{\text{total}}$ & $AOD$ \\ \hline 
        6 & $CC_{\text{low}}$ & $AOD$ & $RH_{\text{middle}}$ & $CC_{\text{low}}$ \\ \hline 
        7 & $AOD$ & $CC_{\text{low}}$ & $CC_{\text{low}}$ & $PW_{\text{total}}$ \\ \hline 
        8 & $DOY$ & $CW_{\text{low}}$ & $AOD$ & $PW_{\text{middle}}$ \\ \hline 
        9 & $PW_{\text{total}}$ & $CC_{\text{high}}$ & $PW_{\text{total}}$ & $CC_{\text{high}}$ \\ \hline 
        10 & $ANG$ & $T_{\text{low}}$ & $CC_{\text{middle}}$ & $\cos{z}$ \\ \hline 
    \end{tabular}
    \caption{The top 10 ranking of the importance of the predictors in the stepwise-based methods.}
    \label{table:ranking_pred_stepwise}
\end{table}

\begin{table}
    \centering
    \begin{tabular}{|c|c|c|c|} \hline
        Rank & QRF & GRF & GBDT \\ \hline
        1 & $G$ & $G$ & $G$ \\ \hline 
        2 & $DIR_{\text{surf}}$ & $DIR_{\text{surf}}$ & $DIR_{\text{surf}}$ \\ \hline 
        3 & $CW_{\text{total}}$ & $CW_{\text{total}}$ & $CW_{\text{total}}$ \\ \hline 
        4 & $CC_{\text{total}}$ & $CC_{\text{total}}$ & $CC_{\text{total}}$ \\ \hline 
        5 & $RH_{\text{low}}$ & $CC_{\text{low}}$ & $RH_{\text{low}}$ \\ \hline 
        6 & $CC_{\text{low}}$ & $RH_{\text{low}}$ & $PW_{\text{total}}$ \\ \hline 
        7 & $CW_{\text{low}}$ & $CW_{\text{low}}$ & $CC_{\text{low}}$ \\ \hline 
        8 & $PW_{\text{total}}$ & $RAIN$ & $ANG$ \\ \hline 
        9 & $RAIN$ & $CW_{\text{middle}}$ & $PW_{\text{low}}$ \\ \hline 
        10 & $CW_{\text{middle}}$ & $PW_{\text{total}}$ & $RH_{\text{middle}}$ \\ \hline 
    \end{tabular}
    \caption{The top 10 ranking of the importance of the predictors in the tree-based methods.}
    \label{table:ranking_pred_tree}
\end{table}

We see that the raw HA global radiation forecast ($G$) is the most important predictor for almost all methods. This is what one would expect, because $G$ already accounts for all kinds of effects from other HA variables. It does however not account optimally for these effects and consequently other predictors are also important, such as direct radiation, relative humidity, cloud cover and cloud water. The relative humidity is more important than the direct radiation in the stepwise-based methods, whereas for the tree-based methods it is the opposite way. This can be explained by the fact that if a stepwise-based method selects global radiation as a predictor, then the direct radiation becomes less useful as a predictor, due to the high correlation between both. In that case the relative humidity becomes more important. For the tree-based methods this effect is less apparent due to the independence between different branches of the trees. Therefore, global radiation can be important in one branch and direct radiation in another branch, and consequently they are both important considering the entire tree. \newline After $G$, the direct radiation and relative humidity, cloud predictors such as $CC_{\text{total}}$ or $CW_{\text{total}}$ follow in most methods. Sometimes the cloud predictors are even higher in the ranking. The aerosol predictors from CAMS are quite important in the stepwise-based methods, with $AOD$ as most important of the three. For the tree-based methods the aerosol predictors are of medium importance seen over all trees, but they can still be useful in certain branches of the trees. The aerosol predictors are in general of lower importance than cloud related predictors. This can be explained by the fact that clouds block on average more incoming radiation than aerosols and are consequently more responsible for the uncertainty in forecast global radiation. For the predictors with multiple layers, the lower component is in general more important than the middle or higher component. Therefore the lower atmosphere has more influence on global radiation. \newline
Furthermore, it is good to note that although some predictors are more important than others, they are all used (seen over all fits) in forecasting global radiation. Also, the tree-based methods have the advantage over the stepwise-based methods that they make use of all predictors in all fits, whereas for the stepwise-based methods we set a limit on the amount of predictors to prevent overfitting.

\subsection{Verification of deterministic forecasts}

\begin{figure}[!ht]
    \centering
    \includegraphics[width = \linewidth]{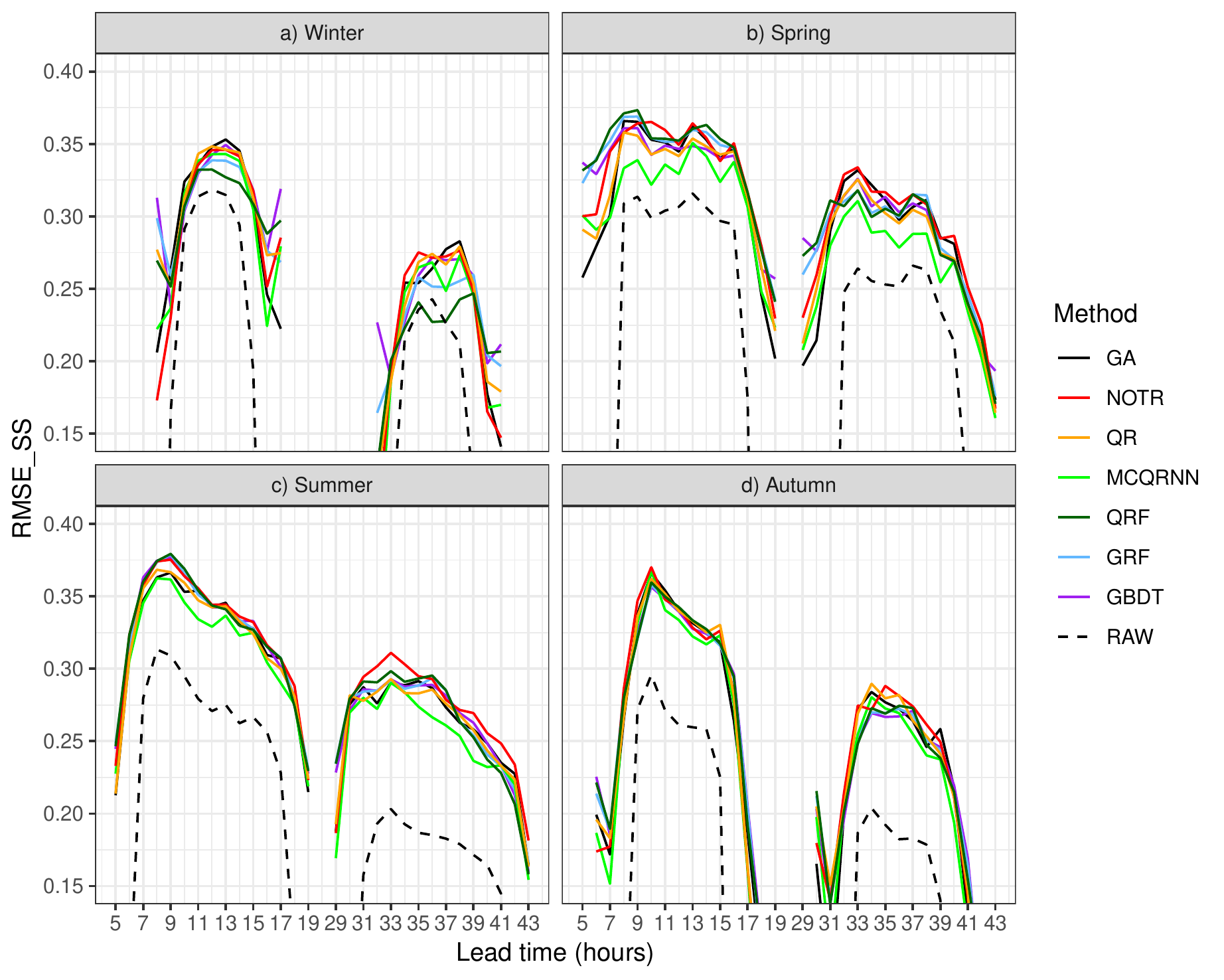}
    \caption{RMSE\_SS versus lead time for the four seasons and averaged over all stations. The abbreviations of the methods are explained in Section \ref{sec:regMethods} and RAW stands for the raw forecast.}
    \label{fig:RMSE_SS_time}
\end{figure}

We computed the RMSE\_SS for all lead times, all stations, the four seasons and the different methods and in Figure \ref{fig:RMSE_SS_time} we plot the RMSE\_SS against the lead time for the four seasons. All methods are more skillful than the raw forecast and have therefore an increased accuracy. The RMSE\_SS with respect to the raw forecast ranges between 2-3\% in winter, 5-7.5\% in spring and 7.5-10\% in summer and fall. For lead times close to the night, the skill drops due to the fact that the forecasts get closer to the sample climatology (because both approach zero) and the latter becomes therefore harder to beat. The second forecast day shows lower RMSE\_SS than the first day as expected. \newline
In winter and spring the RMSE\_SS is relatively stable during the day. Only QRF and MCQRNN perform slightly worse than the other methods in respectively winter and spring. In summer and autumn we see that the skill scores drop during the day, which might be caused by convection (see Figure \ref{fig:timeseriesCSIobs}) as it is difficult to forecast by NWP models. The highest skill scores are reached in the morning. In summer, MCQRNN performs slightly worse than the other methods. In autumn, all methods perform similarly according to the RMSE\_SS. 

\subsection{Probabilistic forecasts for selected cases}

\begin{figure}[!ht]
\centering
    \subfigure[GA method]{\includegraphics[width = 0.48\linewidth]{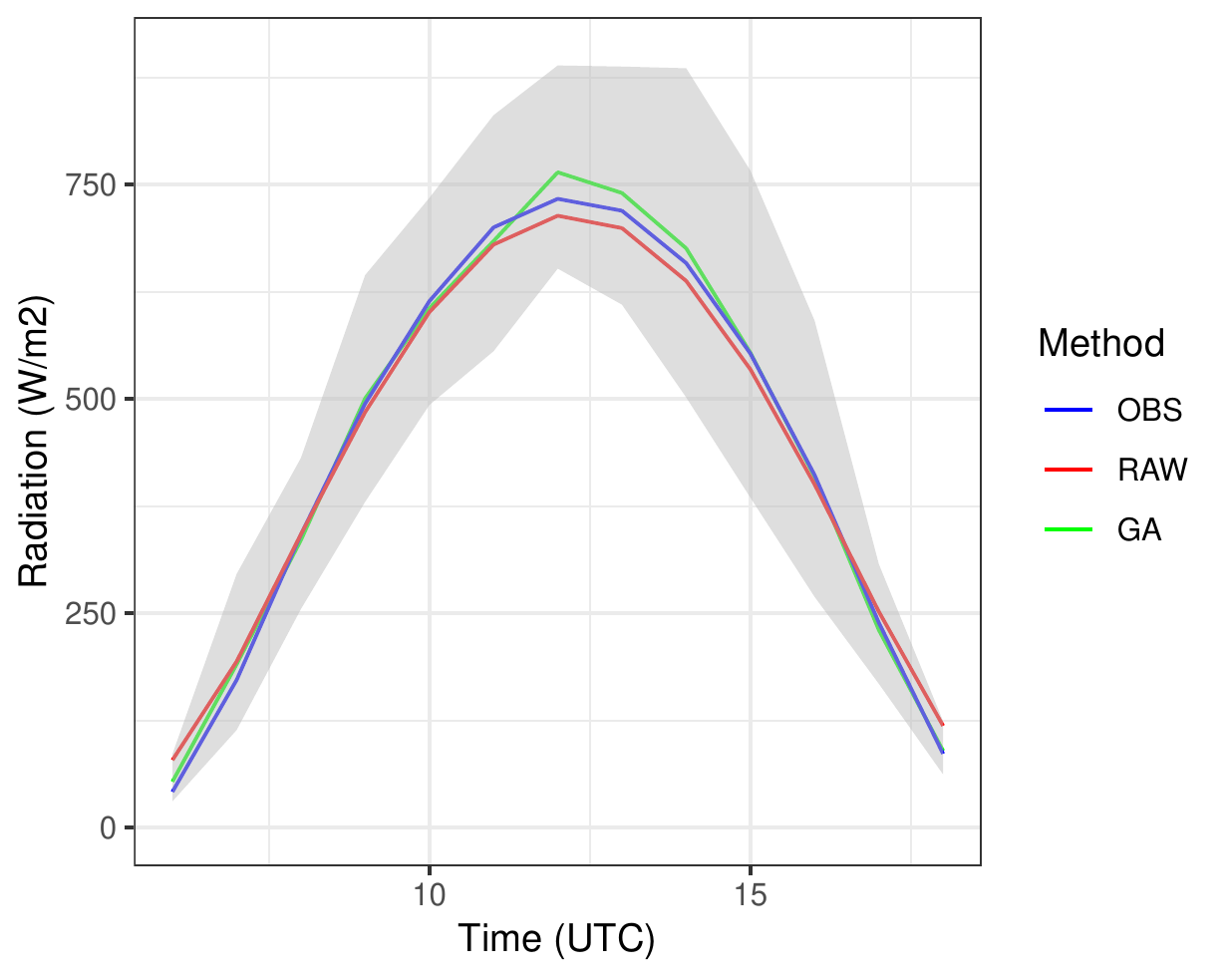}}
    \subfigure[QRF method]{\includegraphics[width = 0.48\linewidth]{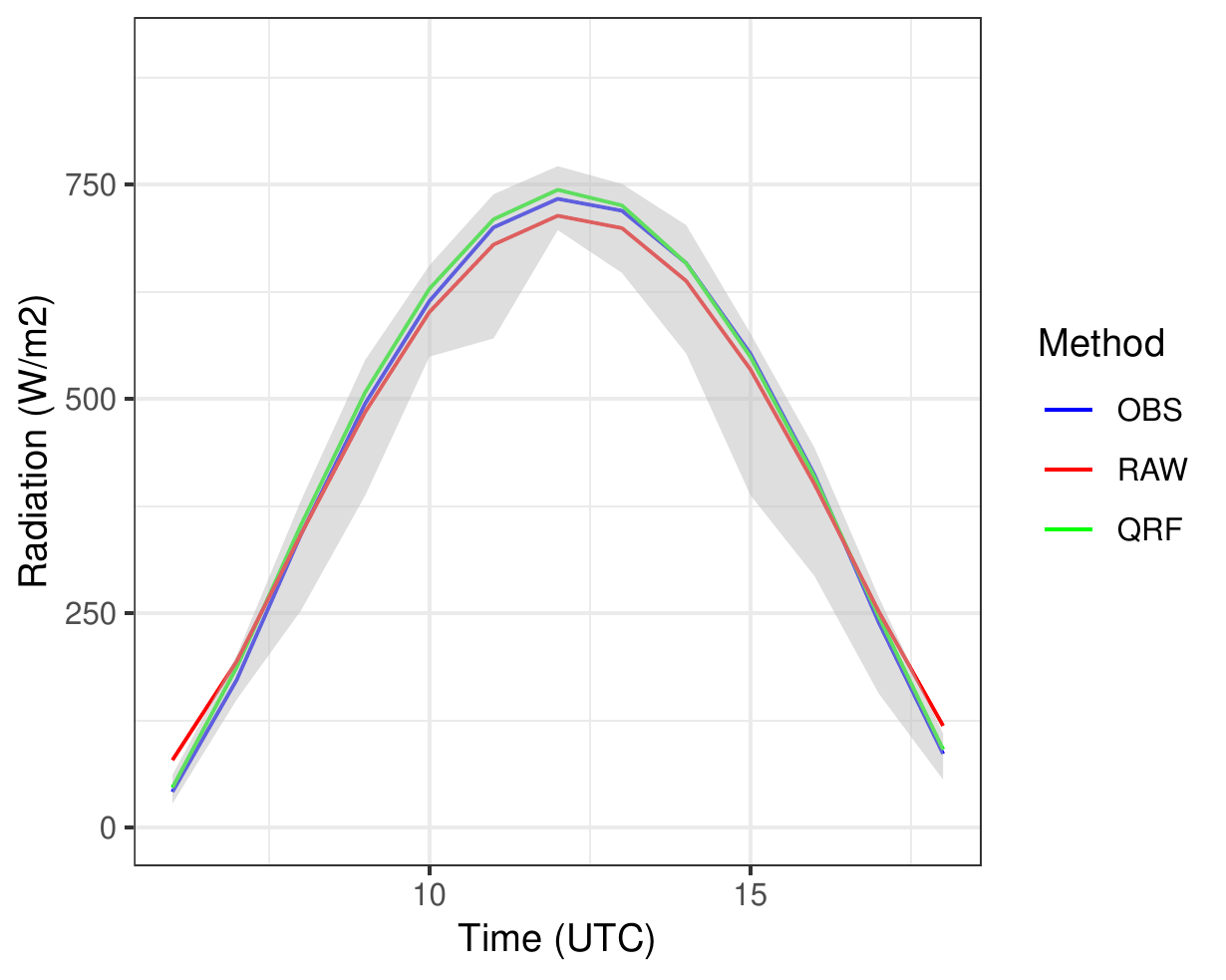}}
    \caption{Clear sky case of April 9, 2017 at station Cabauw (see Figure \ref{fig:spatialCSIobs}). OBS stands for the observations, RAW for the raw forecast, GA for the gamma distribution and QRF for the quantile regression forest. The gray area denotes the probability distribution from the 0.02 until the 0.98 quantile.}
    \label{fig:20170409}
\end{figure}

Three interesting cases are shown for respectively a clear sky, fully overcast and partly overcast day at station Cabauw (see Figure \ref{fig:spatialCSIobs}). First, we look at the case of April 9, 2017, which was a day without clouds. We compare the raw forecast and post-processed forecasts for lead times +6h till +18h with the observations for 6-18 UTC in Figure \ref{fig:20170409}. Two methods are chosen, one parametric (GA) and one non-parametric (QRF), for which the median and the forecast range between the lowest and highest quantile (the gray area) are plotted. We see that the raw forecast is already very accurate and this can be seen on most clear sky days. The raw forecast has a small negative bias around 12 UTC and both GA and QRF correct this in the median, but with a small positive bias, especially for GA. Looking at the uncertainty, we see that QRF is much more certain than GA, with a bandwidth of around 100 W/m$^2$ versus 200 W/m$^2$.  

\begin{figure}[!ht]
\centering
    \subfigure[GA method]{\includegraphics[width = 0.48\linewidth]{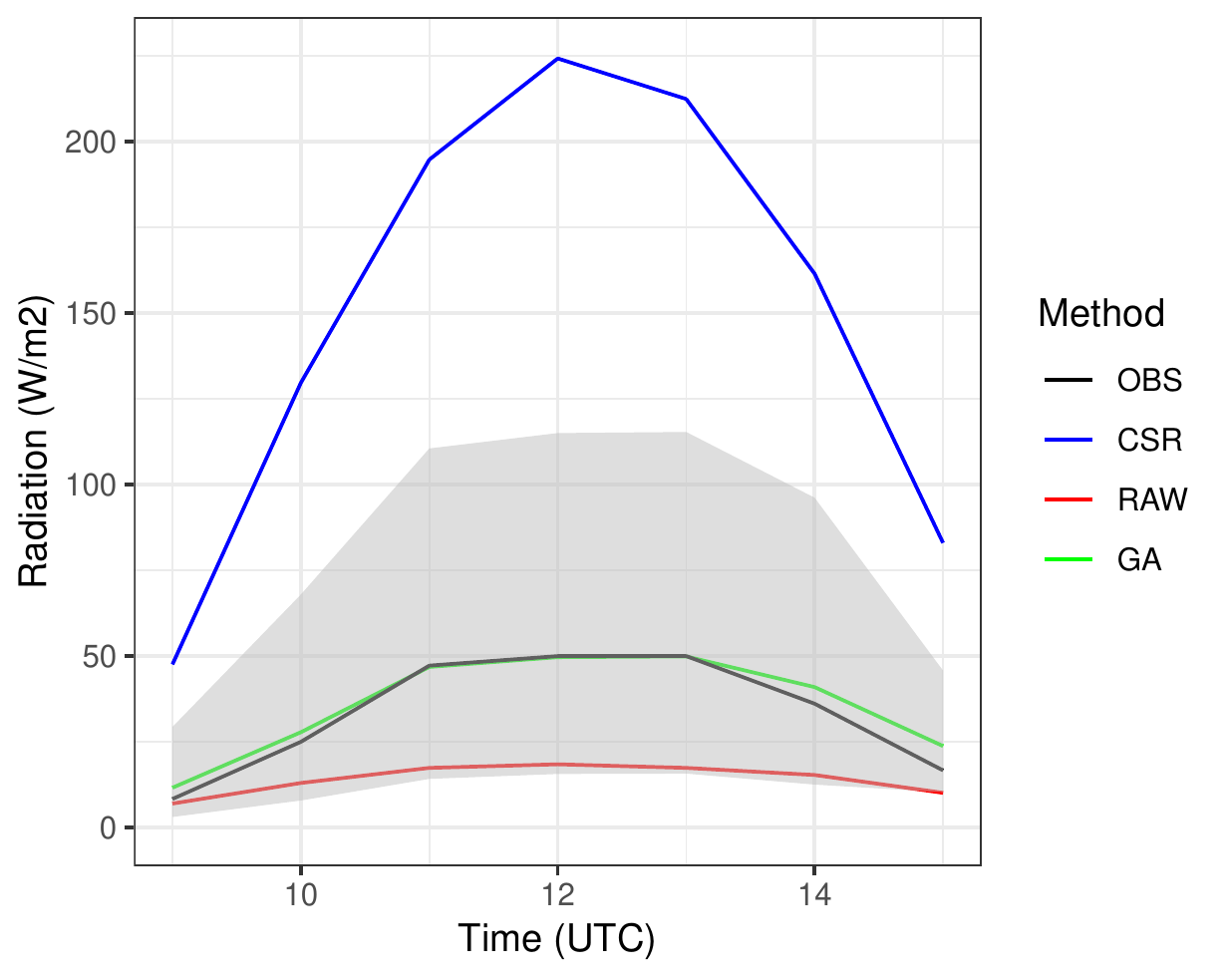}}
    \subfigure[QRF method]{\includegraphics[width = 0.48\linewidth]{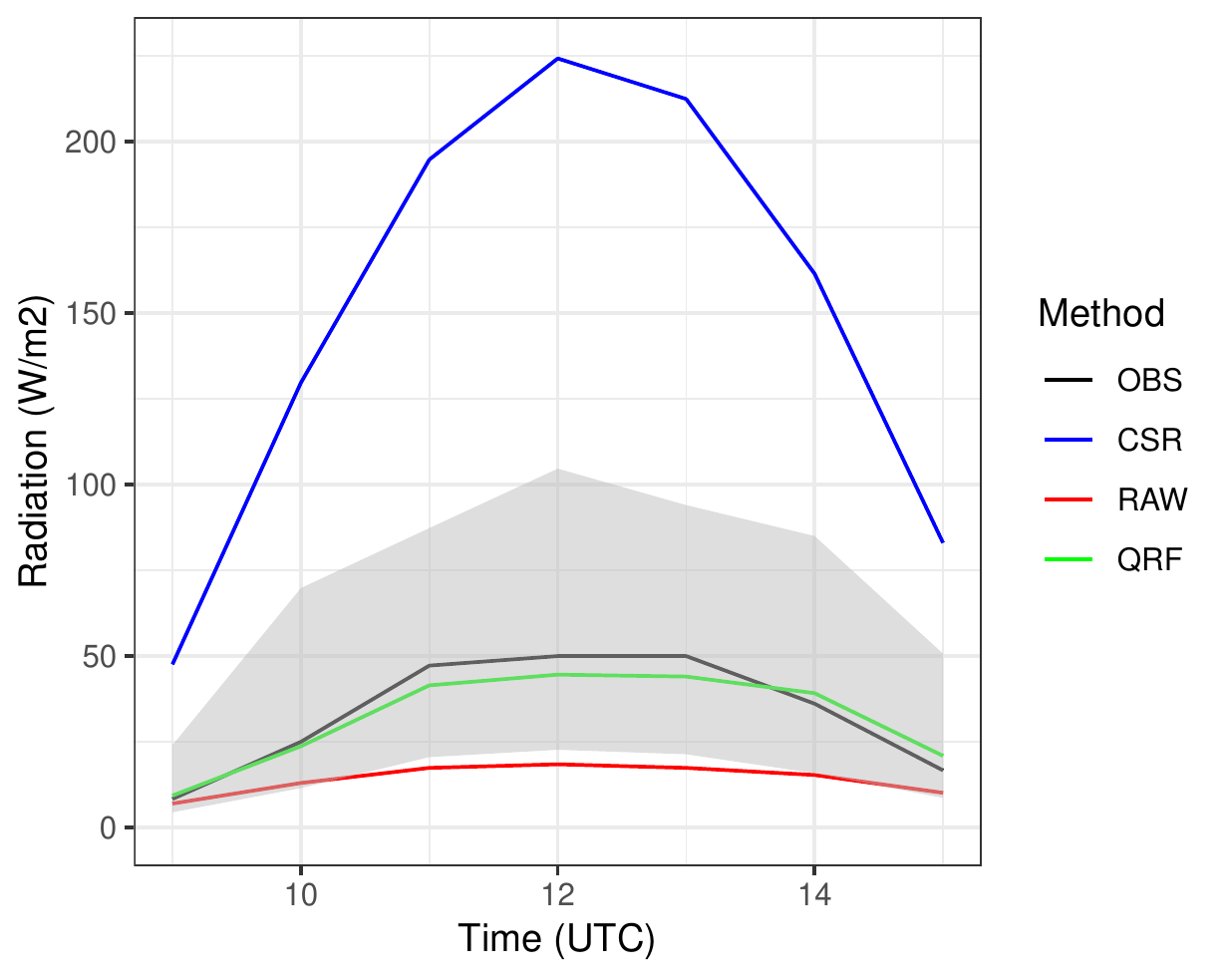}}
    \caption{As Figure \ref{fig:20170409} but for December 31, 2016, which was a fully overcast day. CSR stands for the clear sky radiation.}
    \label{fig:20161231}
\end{figure}

For a fully overcast day we take December 31, 2016. In Figure \ref{fig:20161231} we compare GA and QRF with the raw forecast for lead times +9h till +15h. We see that the raw forecast clearly underestimates the amount of radiation. Both GA and QRF correct for this bias almost completely in the median. QRF generally keeps a small negative bias, while GA has a small positive bias after 13 UTC. Both methods have bandwidths for the uncertainty ranging approximately between 50 and 100 W/m$^2$. This indicates that the methods are very certain that it will be a fully overcast day. GA is slightly more uncertain about the amount of radiation than QRF. The distributions of both GA and QRF are asymmetrical and skewed towards higher (than the median) amounts of radiation, which is (physically) realistic in this case, given that the lower bound of radiation should be strictly greater than zero in the middle of the day. 

\begin{figure}[!ht]
\centering
    \subfigure[GA method]{\includegraphics[width = 0.48\linewidth]{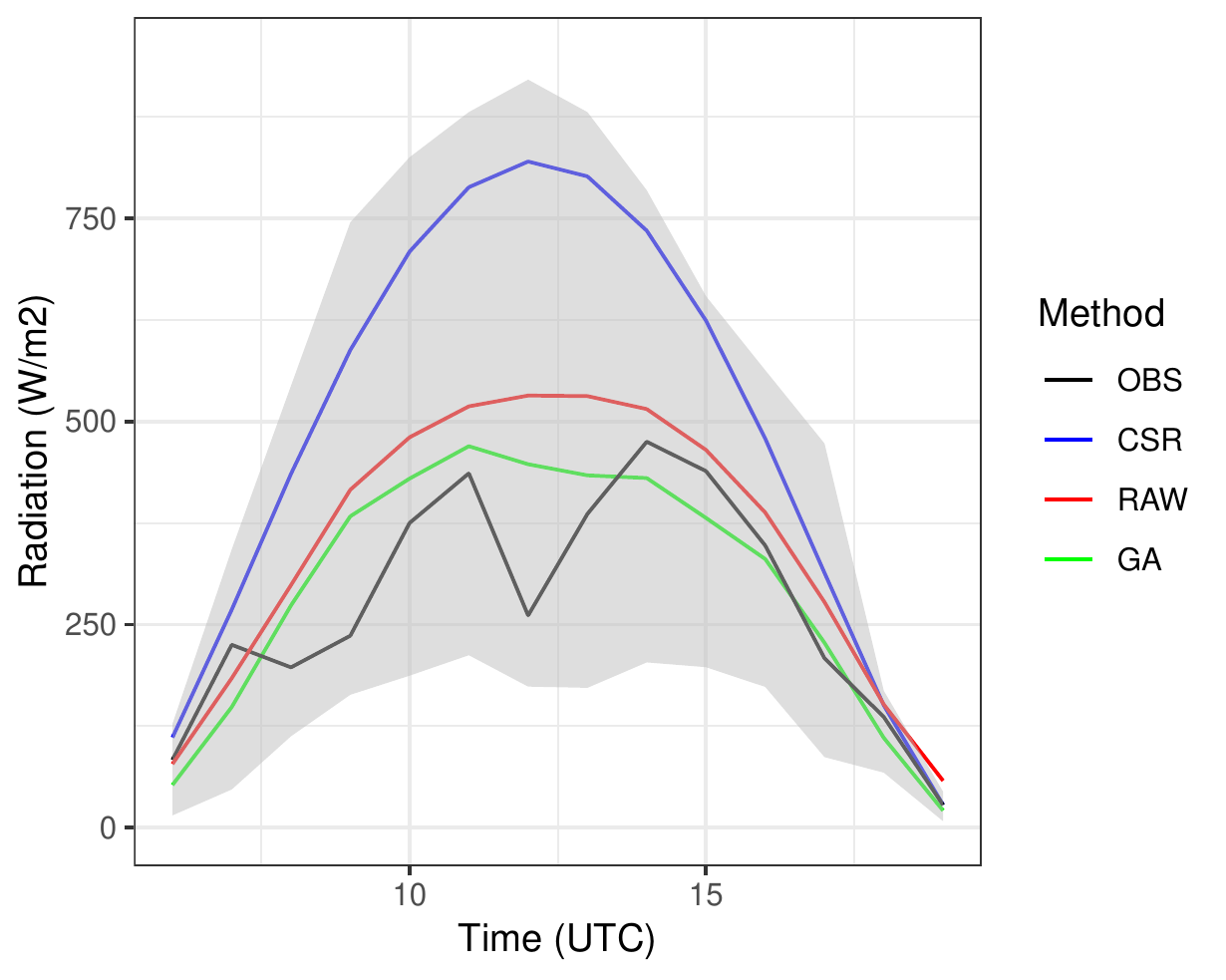}}
    \subfigure[QRF method]{\includegraphics[width = 0.48\linewidth]{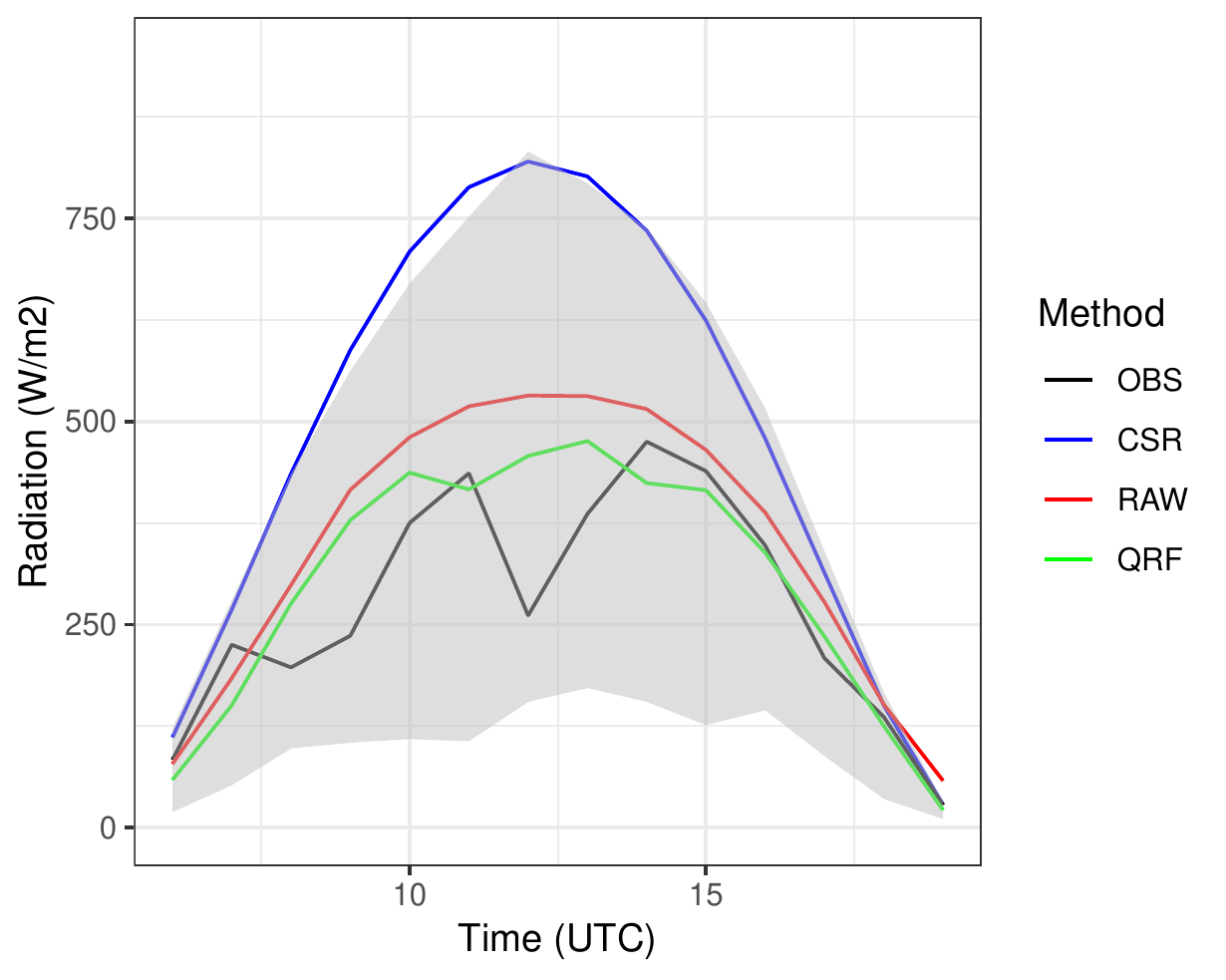}}
    \caption{As Figure \ref{fig:20170409} but for April 23, 2017, which was a partly overcast day. CSR stands for the clear sky radiation.}
    \label{fig:20170423}
\end{figure}

Finally, as partly overcast day we take April 23, 2017. In Figure \ref{fig:20170423} we compare GA and QRF with the raw forecast for lead times +6h till + 19h. In this case the observations are varying more frequently during the day and none of the forecasts are able to forecast these variations. However, the medians of both GA and QRF are closer to the observations than the raw forecast and the median of QRF is better than the median of GA. The uncertainty bandwidths are now indicating that every weather condition between fully overcast and completely sunny is possible. For GA, the uncertainty bandwidth reaches above the clear sky radiation (CSR), due to the assumptions made for that method.

\subsection{Verification of probabilistic forecasts}

For verifying the probabilistic forecasts, we show the results from multiple scoring metrics in the following subsections to compare the performance of the different methods.

\subsubsection{Continuous ranked probability skill score}

\begin{figure}[!ht]
    \centering
    \includegraphics[width = \linewidth]{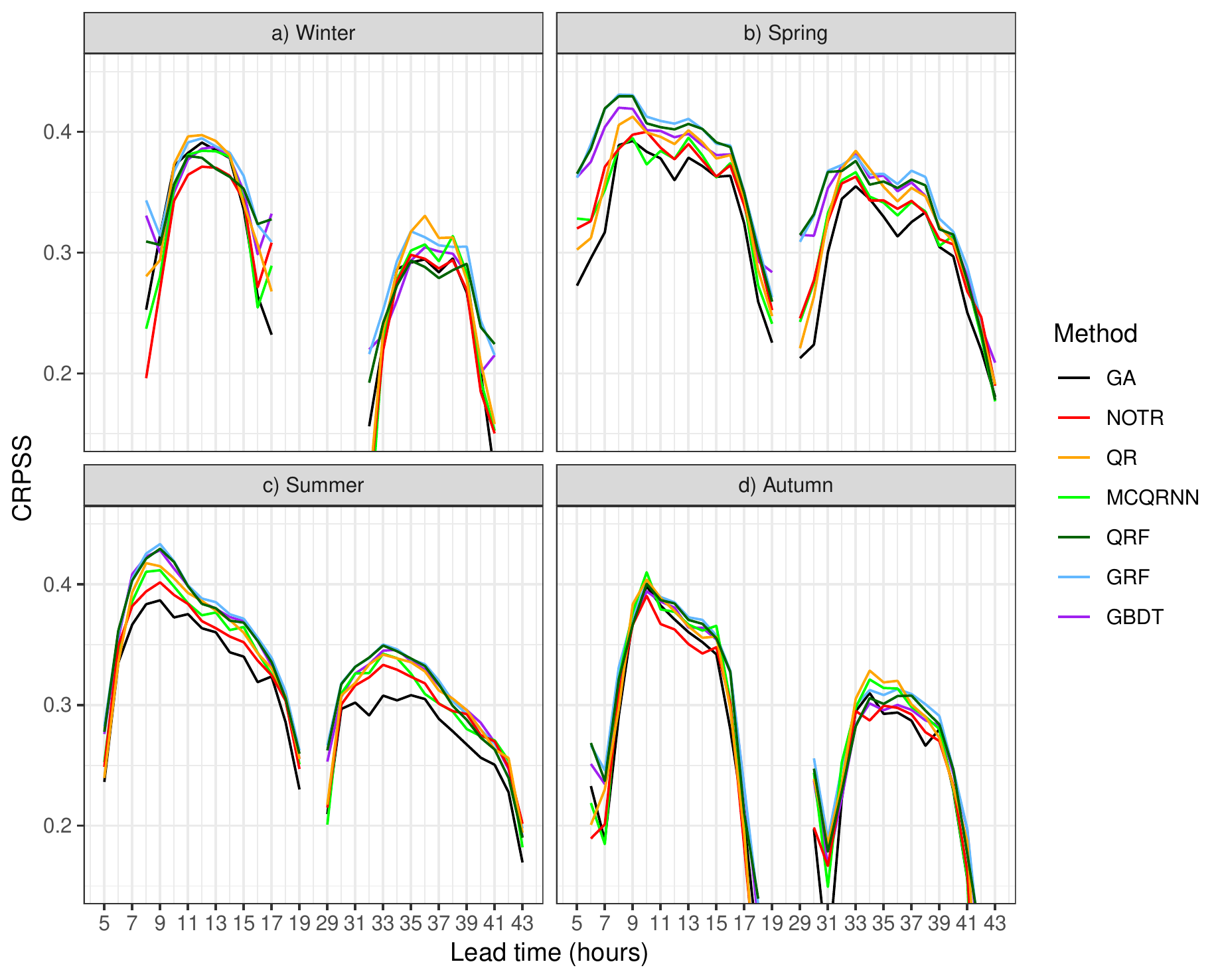}
    \caption{CRPSS versus lead time for the four seasons and averaged over all stations. The abbreviations of the methods are explained in Section \ref{sec:regMethods}.}
    \label{fig:CRPSS_time}
\end{figure}

Starting with the CRPSS, in Figure \ref{fig:CRPSS_time} we plot the CRPSS over the lead times for the different methods in the four seasons. All methods are less skillful closer to the night than during the day due to the fact that the forecasts get closer to the sample climatology (because both approach zero) and the latter becomes therefore harder to beat. The second day shows as expected less skill than the first day. All methods have similar performance with skill scores around 0.4 on day 1 and 0.3-0.35 on day 2. In spring and summer the methods have more skill than in winter and autumn. In winter and spring the skill is relatively stable during the day, while in summer and autumn the skill is best in the morning and decreases during the day. The difficulty in forecasting convection might be the cause of this. In spring and summer, GA performs worse than the other methods, while in winter and autumn the difference in performance between the methods is smaller. Overall, QR and GRF perform best in terms of the average CRPSS. \newline

\begin{figure}[!ht]
    \centering
    \includegraphics[width = \linewidth]{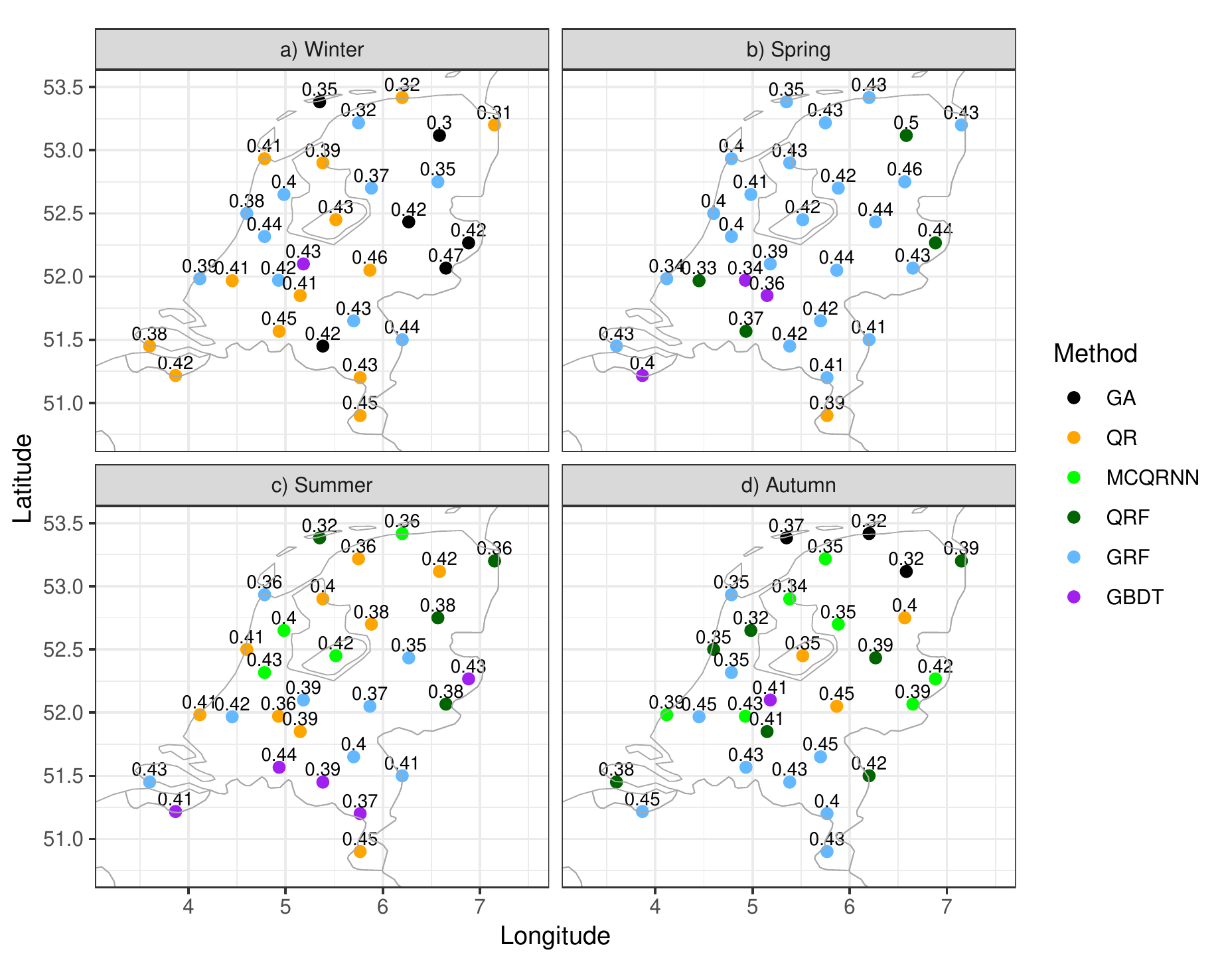}
    \caption{CRPSS for all stations and the four seasons at lead time +12h. Only the maximum value over the methods is shown. The abbreviations of the methods are explained in Section \ref{sec:regMethods}.}
    \label{fig:CRPSS_space}
\end{figure}

It is also informative to look at the spatial patterns of the CRPSS and therefore we plot the maximum value for each station for lead time +12h in Figure \ref{fig:CRPSS_space}. The maxima range between 0.3 and 0.47. We see that the method reaching the maximum varies between the seasons and the stations, only NOTR never performs best. In winter QR is most skillful at most stations, while in spring it is GRF. In summer and autumn it is less clear which method is most skillful overall. Although the skill varies between stations, there is not a clear spatial pattern. There are some seasonal variations, for example in winter and autumn the stations near the coast have mostly less skill than stations more inland. This might occur because convection over sea causes more clouds over stations near the coast.   

\subsubsection{Brier skill score}

\begin{figure}[!ht]
    \centering
    \includegraphics[width = \linewidth]{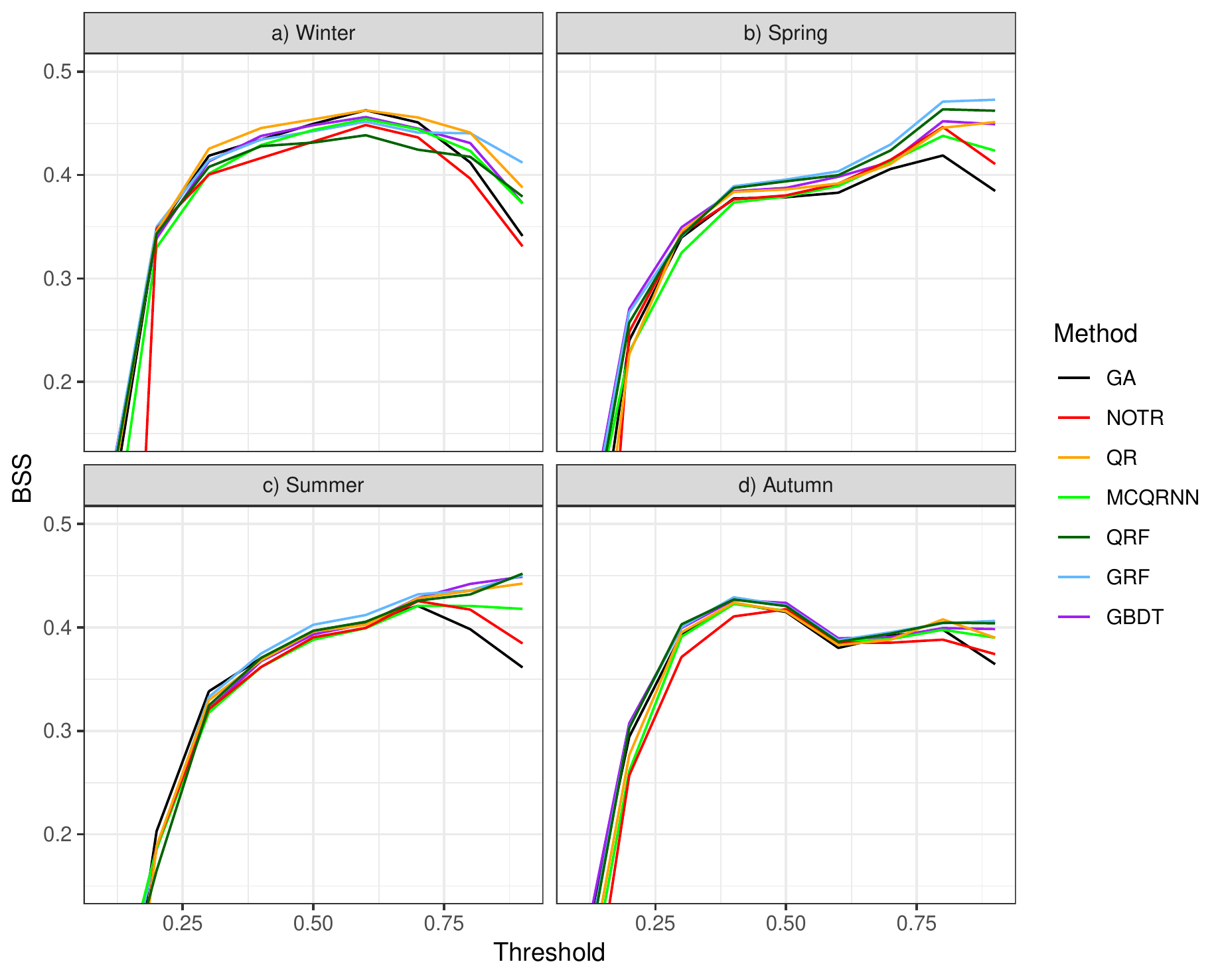}
    \caption{BSS versus CSI threshold at lead time +12h for the four seasons, averaged over all stations. The abbreviations of the methods are explained in Section \ref{sec:regMethods}.}
    \label{fig:BSS}
\end{figure}

To investigate the performance of the methods for different CSI values, we plot in Figure \ref{fig:BSS} the BSS versus the CSI threshold for the four seasons at lead time +12h. The BSS increases as a function of CSI in spring and summer, while in winter it decreases again for higher CSI. In autumn the BSS is stable around 0.4 for CSI values higher than 0.3. The BSS increase in spring and summer is probably caused by the higher predictability of clear sky versus convective situations. While the differences between the methods are small for lower CSI values, it is interesting to note that for thresholds higher than 0.7 (corresponding to (near) clear sky conditions), there is a bifurcation in the skill, especially in spring and summer. The tree-based methods and QR show higher BSS than the other methods. Consequently, the global radiation under (near) clear sky conditions is better forecast by the tree-based methods and QR. For tree-based methods, this can be explained by the fact that trees can create separate branches for clear sky cases. QR fits each quantile separately and can therefore better narrow the distribution in clear sky situations to predict those situations more accurately. 

\subsubsection{Reliability diagrams}

\begin{figure}[!ht]
    \centering
    \includegraphics[width = \linewidth]{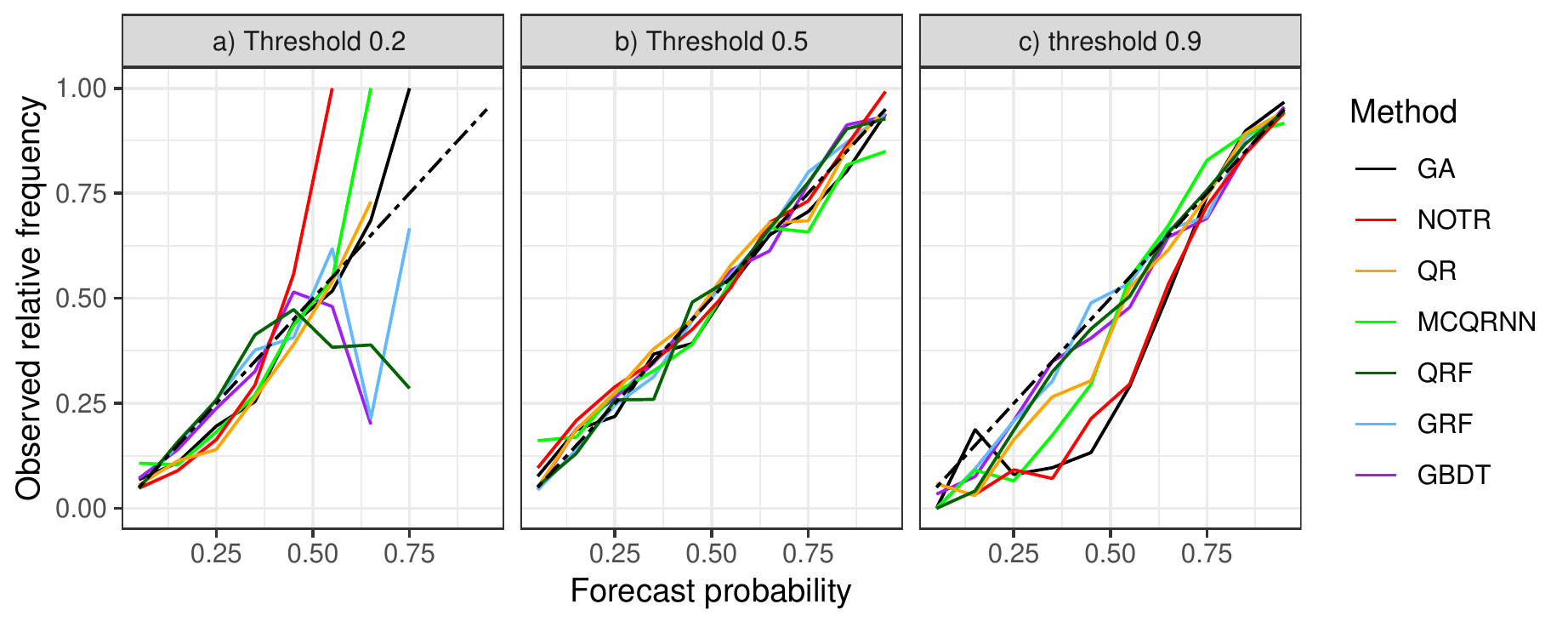}
    \caption{Reliability diagrams for different CSI thresholds at lead time +12h in summer, for all stations together. The abbreviations of the methods are explained in Section \ref{sec:regMethods}.}
    \label{fig:Reliab}
\end{figure}

To access the reliability of the methods, in Figure \ref{fig:Reliab} the reliability diagrams for CSI thresholds 0.2, 0.5 and 0.9 for lead time +12h in summer are shown. For the threshold 0.2 all methods are unreliable for high forecast probabilities, due to a low amount of cases. For the low forecast probabilities the tree-based methods are very reliable and almost on top of the diagonal line. The stepwise-based methods are slightly less reliable. For the threshold 0.5 all methods are very reliable and there is no method that clearly performs better or worse than the rest of the methods. For the threshold 0.9, all methods are reliable for the high forecast probabilities, but for lower forecast probabilities the tree-based methods are again most reliable. The parametric methods are least reliable and QR and MCQRNN are in between. In general the tree-based methods are most reliable, especially under clear sky conditions.

\subsubsection{Potential economic value}

\begin{figure}[!ht]
    \centering
    \includegraphics[width = \linewidth]{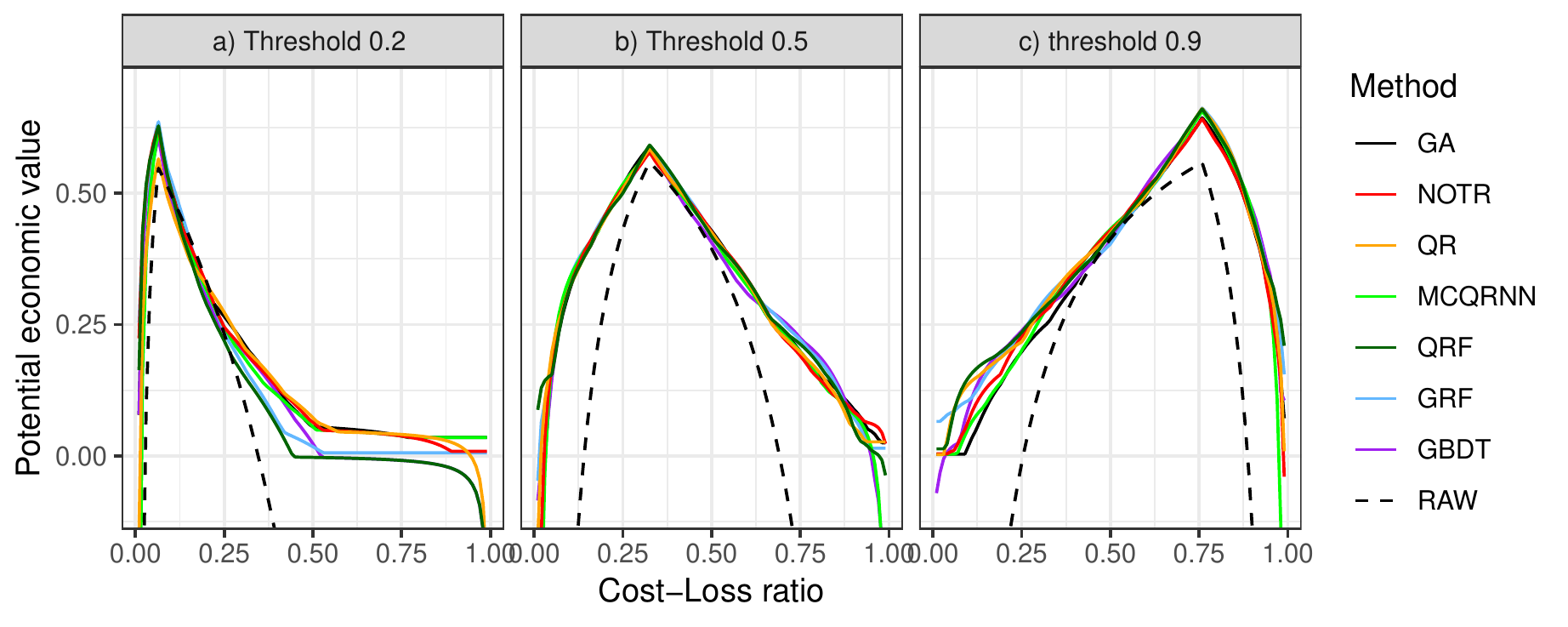}
    \caption{Potential economic value versus cost-loss ratio for different CSI thresholds at lead time +12h in summer, for all stations together. RAW stands for the raw forecast and the rest of the abbreviations are explained in Section \ref{sec:regMethods}.}
    \label{fig:PEV_results}
\end{figure}

To further study the behaviour for different CSI values, in Figure \ref{fig:PEV_results} we plot the potential economic value (PEV) versus the cost-loss ratio for CSI thresholds 0.2, 0.5 and 0.9 at lead time +12h in summer. For all thresholds, the methods show higher PEV values than the raw forecast and show positive PEV values over a wider cost-loss ratio range (the full range between 0 and 1 for thresholds 0.5 and 0.9). For threshold 0.2, the tree-based methods perform worse than the stepwise-based methods for cost-loss ratios above 0.4. For the threshold 0.5, all methods have equal performance. For the threshold 0.9, the tree-based methods and QR perform slightly better than the other methods for cost-loss ratios below 0.2.

\section{Discussion}{\label{sec:discussion}}

The performance of the methods presented here depends on the lead time, the station location and the season. Nevertheless, in general all methods had similar performance. It also depends on the scoring metrics, that focus on different aspects, like accuracy and reliability, and on deterministic or probabilistic forecasts. We often saw that quantile regression and generalized random forests performed best and that the tree-based methods outperformed the stepwise-based methods in (near) clear sky conditions. In the discussion presented here we compare our results with results from other papers, where comparing methods is the primary goal, and look for differences and similarities. \newline
\citet{David_2018} made a comparison between probabilistic forecasts using random forests, neural networks, (weighted) quantile regression, gradient boosting decision trees, recursive GARCH and the sieve bootstrap. They used endogenous data consisting of past observations of global radiation and CSI as predictors for forecasting CSI. They came to the conclusion that all methods performed very similarly, in agreement with our results. They based this on verification measures including the CRPS, reliability diagrams and rank histograms. They found (weighted) quantile regression and gradient boosting decision trees as most efficient methods for generating probabilistic forecasts in terms of complexity and computational time. They concluded that neural networks are the least efficient with a large computational time and requiring sufficient scientific background for good performance. \newline
\citet{Mohammed_2016} made a comparison between deterministic forecasts using decision trees, nearest neighbours, gradient boosting decision trees, random forests and both lasso and ridge regression. They used NWP output from the ECMWF as predictors to forecast solar power production. They found gradient boosting decision trees as best performing in terms of the RMSE and mean absolute error (MAE). In our study, gradient boosting decision trees performed well, but not the best, according to the RMSE. \newline
\citet{Zamo_2014} made a comparison between probabilistic forecasts using quantile regression and quantile regression forests. They used NWP output from the PEARP model at M\'et\'eo France. They forecast solar power production and verify the forecasts using the CRPS and rank histograms. They found that the best performing method depends on the specific settings and concluded that quantile regression and quantile regression forests performed similar in general, which is in agreement with our results. \newline
Lastly, \citet{Voyant_2018} made a comparison between probabilistic forecasts using quantile regression forests and gradient boosting decision trees with past CSI observations as predictor set. They generate prediction intervals for the CSI and verify them using the gamma test, which is based on the mean interval length and prediction interval coverage probability. The authors found that gradient boosting decision trees performed best. In our study, quantile regression forests and gradient boosting decision trees performed almost similar.

\section{Conclusions}{\label{sec:conclusion}}

In this paper we compared 7 regression methods. Using these methods, we produced probabilistic forecasts of global radiation with the raw forecast (the deterministic global radiation forecast from HA) together with other relevant meteorological variables as potential predictors. We saw that all potential predictors were used, and that global radiation, direct radiation, relative humidity, cloud cover and cloud water were most important. The aerosol predictors from CAMS were less important than the radiation and cloud-based predictors from HA, but they were still often selected, particularly the aerosol optical depth in the stepwise-based methods. Aside from that, the tree-based methods have the advantage of using all predictors in all fits, whereas for the stepwise-based methods the amount of predictors used in a fit is limited. We verified the forecasts in a deterministic way using the root mean squared error skill score. All methods had an improvement between 2\% and 10\% upon the raw forecast, depending on the season and lead time. In visualizing the probabilistic information in the forecasts we looked at three interesting cases with different weather conditions: a clear sky, fully overcast and partly overcast day. In all cases the medians of the forecasts were close to the observations, but the uncertainty in forecast radiation differed between the methods (mostly on the clear sky and partly overcast day). Next, we verified the methods in a probabilistic way using the continuous ranked probability skill score, Brier skill score, reliability diagrams and the potential economic value. We saw that the performance of the methods depended on the time of the day, the forecast lead time, the geographical location, the season and the weather situation (clear sky or cloudy). In general all methods performed similarly, but depending on the verification measure there are some methods that performed better than others. According to the continuous ranked probability skill score, quantile regression and generalized random forests performed slightly better and the gamma and truncated normal distributions slightly worse than the other methods. For the Brier skill score there was not a clear best performing method, but we saw that clear sky conditions are better forecast by tree-based methods than stepwise-based methods. Also the tree-based methods are more reliable, according to the reliability diagrams. Next, the probabilistic forecasts were compared with the deterministic raw forecast using the potential economic value to show the added value of uncertainty information in the probabilistic forecasts. All methods achieved both higher potential economic values and positive potential economic values over a wider cost-loss ratio range than the raw forecast. Therefore the probabilistic forecasts are very useful for users of solar energy. They can make better decisions based on the improved accuracy and information about the uncertainty. \newline 
Future research could focus on more complex economic models for making a comparison between the methods. Another research direction is to investigate other distributions than the ones tried in this paper or other non-parametric regression methods. Furthermore, the effect of more potential predictors (for example diffuse radiation) could be investigated. Also predictors taken from the (relatively new) HA ensemble prediction system could be used, such as the ensemble mean or ensemble variance of the global radiation forecast, and would likely be better predictors than those from the deterministic HA forecasts. \newline
The methods presented here are all based on techniques of statistical post-processing. A different, physical approach is to improve the NWP models to lead to better predictors. This could be for example improving the representation of clouds and aerosols in the model. This would also improve the forecasts of global radiation. However, statistical post-processing is still expected to improve skill of those forecasts. 

\section*{Acknowledgements}

We thank Jan Fokke Meirink (KNMI) for providing the aerosol forecasts from CAMS, consisting of the aerosol optical depth, \r{A}ngstr\"{o}m exponent and ozone. We thank Jason Frank (Utrecht University), Wim de Rooy, Jan Barkmeijer, Sander Tijm and Sibbo van der Veen (all from KNMI) for their help in the study described in this paper. \\

This research did not receive any specific grant from funding agencies in the public, commercial, or not-for-profit sectors. \\

Declarations of interest: none

%\section*{References}

%\nocite{*}
\bibliographystyle{elsarticle-harv} 
\bibliography{Refs}

\end{document}